\documentclass[aps,prd,twocolumn,showpacs,nofootinbib]{revtex4-1}

\usepackage{amssymb} \usepackage{amsmath} \usepackage{graphicx}
\usepackage{epsfig,latexsym}
\usepackage{slashed}
\usepackage{mathtools}

\RequirePackage{xspace} \allowdisplaybreaks

\begin{document}

\def\bef{\begin{figure}}
\def\eef{\end{figure}}

\newcommand{\nl}{\nonumber\\}

\newcommand{\ans}{ansatz }
\newcommand{\be}[1]{\begin{equation}\label{#1}}
\newcommand{\beq}{\begin{equation}}
\newcommand{\ee}{\end{equation}}
\newcommand{\beqn}[1]{\begin{eqnarray}\label{#1}}
\newcommand{\eeqn}{\end{eqnarray}}
\newcommand{\bd}{\begin{displaymath}}
\newcommand{\ed}{\end{displaymath}}
\newcommand{\mat}[4]{\left(\begin{array}{cc}{#1}&{#2}\\{#3}&{#4}
\end{array}\right)}
\newcommand{\matr}[9]{\left(\begin{array}{ccc}{#1}&{#2}&{#3}\\
{#4}&{#5}&{#6}\\{#7}&{#8}&{#9}\end{array}\right)}
\newcommand{\matrr}[6]{\left(\begin{array}{cc}{#1}&{#2}\\
{#3}&{#4}\\{#5}&{#6}\end{array}\right)}
\newcommand{\cvb}[3]{#1^{#2}_{#3}}
\def\lsim{\raise0.3ex\hbox{$\;<$\kern-0.75em\raise-1.1ex
e\hbox{$\sim\;$}}}
\def\gsim{\raise0.3ex\hbox{$\;>$\kern-0.75em\raise-1.1ex
\hbox{$\sim\;$}}}
\def\abs#1{\left| #1\right|}
\def\simlt{\mathrel{\lower2.5pt\vbox{\lineskip=0pt\baselineskip=0pt
           \hbox{$<$}\hbox{$\sim$}}}}
\def\simgt{\mathrel{\lower2.5pt\vbox{\lineskip=0pt\baselineskip=0pt
           \hbox{$>$}\hbox{$\sim$}}}}
\def\unity{{\hbox{1\kern-.8mm l}}}
\newcommand{\eps}{\varepsilon}
\def\ep{\epsilon}
\def\ga{\gamma}
\def\Ga{\Gamma}
\def\om{\omega}
\def\omp{{\omega^\prime}}
\def\Om{\Omega}
\def\la{\lambda}
\def\La{\Lambda}
\def\al{\alpha}
\newcommand{\ov}{\overline}
\renewcommand{\to}{\rightarrow}
\renewcommand{\vec}[1]{\mathbf{#1}}
\newcommand{\vect}[1]{\mbox{\boldmath$#1$}}
\def\tm{{\widetilde{m}}}
\def\mcirc{{\stackrel{o}{m}}}
\newcommand{\Dm}{\Delta m}
\newcommand{\dm}{\varepsilon}
\newcommand{\tanb}{\tan\beta}
\newcommand{\nbar}{\tilde{n}}
\newcommand\PM[1]{\begin{pmatrix}#1\end{pmatrix}}
\newcommand{\up}{\uparrow}
\newcommand{\down}{\downarrow}
\def\omE{\omega_{\rm Ter}}

%

\newcommand{\Dsusy}{{susy \hspace{-9.4pt} \slash}\;}
\newcommand{\DCP}{{CP \hspace{-7.4pt} \slash}\;}
\newcommand{\mc}{\mathcal}
\newcommand{\gr}{\mathbf}
\renewcommand{\to}{\rightarrow}
\newcommand{\gtc}{\mathfrak}
\newcommand{\wh}{\widehat}
\newcommand{\br}{\langle}
\newcommand{\kt}{\rangle}


\def\lsim{\mathrel{\mathop  {\hbox{\lower0.5ex\hbox{$\sim$}
\kern-0.8em\lower-0.7ex\hbox{$<$}}}}}
\def\gsim{\mathrel{\mathop  {\hbox{\lower0.5ex\hbox{$\sim$}
\kern-0.8em\lower-0.7ex\hbox{$>$}}}}}

\def\nn{\\  \nonumber}
\def\de{\partial}
\def\brf{{\mathbf f}}
\def\bbf{\bar{\bf f}}
\def\bF{{\bf F}}
\def\bbF{\bar{\bf F}}
\def\bA{{\mathbf A}}
\def\bB{{\mathbf B}}
\def\bG{{\mathbf G}}
\def\bI{{\mathbf I}}
\def\bM{{\mathbf M}}
\def\bY{{\mathbf Y}}
\def\bX{{\mathbf X}}
\def\bS{{\mathbf S}}
\def\bb{{\mathbf b}}
\def\bh{{\mathbf h}}
\def\bg{{\mathbf g}}
\def\bla{{\mathbf \la}}
\def\bmu{\mathbf m }
\def\by{{\mathbf y}}
\def\bmu{\mbox{\boldmath $\mu$} }
\def\bsig{\mbox{\boldmath $\sigma$} }
\def\bunity{{\mathbf 1}}
\def\cA{{\cal A}}
\def\cB{{\cal B}}
\def\cC{{\cal C}}
\def\cD{{\cal D}}
\def\cF{{\cal F}}
\def\cG{{\cal G}}
\def\cH{{\cal H}}
\def\cI{{\cal I}}
\def\cL{{\cal L}}
\def\cN{{\cal N}}
\def\cM{{\cal M}}
\def\cO{{\cal O}}
\def\cR{{\cal R}}
\def\cS{{\cal S}}
\def\cT{{\cal T}}
\def\eV{{\rm eV}}

%

\title{Dark Matter and Inflation in $R+\zeta R^{2}$ Supergravity}

\author{Andrea Addazi$^1$}\email{andrea.addazi@infn.lngs.it}
\affiliation{$^1$ Dipartimento di Fisica,
 Universit\`a di L'Aquila, 67010 Coppito AQ and
LNGS, Laboratori Nazionali del Gran Sasso, 67010 Assergi AQ, Italy}

\author{Maxim Yu. Khlopov$^2$}\email{khlopov@apc.in2p3.fr}
\affiliation{$^2$ Centre for Cosmoparticle Physics Cosmion;
National Research Nuclear University MEPHI (Moscow Engineering Physics Institute), Kashirskoe Sh., 31, Moscow 115409, Russia;
APC laboratory 10, rue Alice Domon et L\'eonie Duquet 75205 Paris Cedex 13, France}

\begin{abstract}

As is well known, the gravitational degrees of freedom contained in $R+\zeta R^{2}$ (super)gravity
lead to  Starobinsky's potential, in a one-field setting for inflationary
Cosmology that appears favored  by Planck data. In this letter we discuss
another interesting aspect of this model, related to gravitino production,
with emphasis on the corresponding mass spectrum.
Assuming that supersymmetry is broken at a very high scale, 
Super Heavy Gravitino Dark Matter (SHGDM) and Starobinsky's inflation
can be coherently unified in a $R+\zeta R^{2}$ supergravity. 
Gravitinos are assumed to be the Lightest Supersymmetric Particles (LSP)
and are non-thermally produced during inflation, 
in turn originated by a scalar with a Starobinsky's potential. 
Gravitino mass runs with the inflaton field, so that
a continuos spectrum of superheavy gravitinos emerges. 
The theory is implemented with a $U(1)_{R}$ gauge symmetry.
However, in a string UV completion, $U(1)_{R}$-symmetry can be broken 
 by non-perturbative string instantons,  while for consistency of our scenario $U(1)_{R}$ gauge symmetry breaking must be broken in order to 
 generate a soft mass terms for the gravitino and gauginos. 
  R-parity violating operators can be generated at non-perturbative level. 
Gravitinos can decay into very energetic neutrinos and photons 
in cosmological time scale, with intriguing implications for high energy cosmic rays experiments.

\end{abstract}

\maketitle
\section{Introduction}

Starobinsky's model is the simplest $f(R)$-extension 
of the Einstein-Hilbert action \cite{S1}. 
As is well known, 
Starobinsky's $R+\zeta R^{2}$ is a good theory of inflation, 
in agreement with recent Planck data \cite{Ade:2015lrj}. 
Starobinsky's model can be conformally transformed into 
a scalar-tensor theory, where the scalaron has 
an inflation slow-roll potential. This motivated theoretical researches of a supergravity  
embedding the Starobinsky's model. The simplest proposal suggested in Ref.\cite{9,11} entails
a tachyonic instability of the Goldstino 
at large values of the inflaton. 
On the other hand, this problem was solved in Ref.\cite{5,6}
and in Ref.\cite{18a,18b,18c,18d,18e,21,22,Ferrara:2013rsa,Ferrara:2013pla,Ferrara:2014cca,Ferrara:2015ela,Ozkan:2014cua,Ferrara:2014kva,Antoniadis:2014oya,Dudas:2015eha,Ferrara:2015gta,Ferrara:2016buf}
in frameworks of no-scale and Volkov-Akulov supersymmetry. 
As a result, a consistent $R+\zeta R^{2}$ supergravity 
can be obtained without unstable moduli fields. 
This is a necessary (but not sufficient) condition 
for a UV completion in string theories, 
where many other scalar moduli of the compactification are inevitably introduced. 
However, new implications of $R+\zeta R^{2}$ supergravity 
for the non-supersymmetric Starobinsky's model were not fully addressed in literature. 
For instance, possible implications of this model on  
dark matter production were not been analyzed in all the details. 

In this paper we study $R+\zeta R^{2}$ supergravity 
 with local supersymmetry broken 
at scales higher than the inflaton reheating. 
As is well known, a supergravity theory has to contain at least a new
supersymmetric spin $3/2$ partner for the graviton, 
the gravitino. If SUSY is broken at high scales, the gravitino will naturally get
a large soft mass term comparable to the SUSY scale. 
The gauge R-parity symmetry protects the gravitino against R-parity violating couplings. 
The gravitino could decay in R-preserving transitions into other Supersymmetric Standard Model (or Beyond SM) particles.
However, we assume that gravitino is the Lightest Supersymmetric Particle (LSP) of the supersymmetric spectrum. 
On the other hand, $U(1)_{R}$ symmetry cannot be preserved at lower energies, otherwise 
 gauginos cannot have soft mass terms larger than the gravitinos mass. 
We shall comment on possible St\"uckelberg mechanism for $U(1)_{R}$ breaking, associated 
to the presence of non-perturbative effects like exotic D-brane instantons in open strings theories. 
These aspects were recently discussed in the context of intersecting D-brane models
and quiver theories, where exotic instantons can generate $B-L$ violating couplings
with intriguing implications for rare process and baryogenesis
  \cite{Blu1,Ibanez1,Ibanez2,Ibanez3,Blu2,Addazi:2014ila,Addazi:2015ata,Addazi:2015rwa,Addazi:2015hka,Addazi:2015ewa,Addazi:2015yna,Addazi:2015eca,Addazi:2016mtn,Addazi:2015fua,Addazi:2015oba,Addazi:2015goa,Addazi:2016xuh}.
In this scenario, the gravitino problem is avoided: the gravitino is a supermassive particle which does not decay in a time $t\leq  10 \, {\rm s} \div 20\, {\rm min}$, 
so that Big Bang Nucleosynthesis is not affected and ruined
\cite{Khlopov:1984pf,Khlopov:1993ye,Khlopov:2015oda}. 
We will show how a super-heavy gravitino can be non-thermally produced during Starobinsky's inflation, 
 with the right Cold Dark Matter abundance.
On the other hand thermal production of gravitinos will be suppressed 
 if the gravitino is heavier than the inflaton mass
\cite{Rychkov:2007uq}. 
As a result, Cold Dark Matter  is connected to the parameter space of inflation in a unified minimal framework. 
Finally, we shall discuss how gravitinos can be destabilized in cosmological time scakes, 
decaying into very high energy neutrinos, which could be detectable in principle 
by high energy cosmic rays experiments: AUGER, Telescope Array, ANTARES and IceCube.

\section{$R+\zeta R^{2}$ Supergravity }

Let us consider the Lagrangian of $R+\zeta R^{2}$ supergravity coupled to matter
\cite{Ferrara:2013rsa,Ferrara:2013pla,Ferrara:2014cca,Ferrara:2015ela,Antoniadis:2014oya}
\be{L}
\mathcal{L}=-[- L\mathcal{V}_{R}+L\Phi(z,\bar{z})]_{D}+\zeta[\mathcal{W}_{\alpha}(\mathcal{V}_{R})\mathcal{W}^{\alpha}(\mathcal{V}_{R})]
\ee
$$\mathcal{V}_{R}=\log \frac{L}{\mathcal{S}\mathcal{\bar{S}}}$$
where the first term contains the standard Einstein-Hillbert action
while the $R^{2}$ term is generated by the kinetic term of the real superfield $\mathcal{V}_{R}$, 
$\mathcal{S}$ is the compensator field of old minimal supergravity, 
the functional $\Phi(z,\bar{z})$ is the K\"ahler potential of the $z$ scalar fields and 
$L$ is the linear multiplet. 
Eq.(\ref{L}) can be rewritten as 
\be{Unconstrained}
\zeta \mathcal{W}_{\alpha}(\mathcal{U})\mathcal{W}^{\alpha}(\mathcal{U})-\left[\mathcal{S}_{0}\mathcal{\bar{S}}_{0}e^{\mathcal{U}-\mathcal{T}-\mathcal{\bar{T}}}\left(\mathcal{U}+\frac{1}{3}\Phi-\mathcal{T}-\mathcal{\bar{T}}\right) \right]+c.c. 
\ee
where $\mathcal{S}=\mathcal{S}_{0}e^{-\mathcal{T}}$ and $\mathcal{T}+\mathcal{\bar{T}}$ are lagrangian multiplier fields 
which allow to consider an unconstrained vector multiplet $\mathcal{U}$.
The gauged R-symmetry can be implemented as 
\be{V}
\mathcal{V}_{R}\rightarrow \mathcal{V}_{R}+\Omega+\bar{\Omega},\,\,\,z_{I}\rightarrow e^{q_{I}\Omega}z_{I},\,\,\,\mathcal{S}\rightarrow e^{-\Omega}\mathcal{S}
\ee
with $\Omega$ chiral superfield. 
A superpotential compatible with R-symmetry can be included 
as a term $\mathcal{S}_{0}^{3}e^{-3T}\mathcal{W}(z)$ in the lagrangian (\ref{Unconstrained}).

The scalar potential is a sum of $V_{F}$ and $V_{D}$:
\be{VF}
V_{F}=e^{\mathcal{G}}(\mathcal{G}_{A}\mathcal{G}^{AB}\mathcal{G}_{B}-3),\,\,\,\mathcal{G}_{A}=\frac{\partial \mathcal{G}}{\partial Z^{A}},\,\,\,Z^{A}=(T,z^{I})
\ee
$$\mathcal{G}=K+{\rm log}(e^{-3T}\mathcal{W})+{\rm log}(e^{-3\bar{T}}\mathcal{\bar{W}})$$
\be{VD}
2\zeta V_{D}=2\mathcal{G}_{T}+\sum [q_{I}z^{I}\mathcal{G}_{I}+q_{I}\bar{z}^{I}\mathcal{G}_{I}]
\ee
The old Starobinsky's inflaton potential can be recovered 
assuming that 
the F-term is so steep to rapidly drive all $z^{I}$ fields to $\mathcal{W}_{I}=\partial \mathcal{W}/\partial z^{I}\rightarrow 0$,
which is an R-symmetric vacuum. The only contribution to the 
potential comes from the D-term: 
\be{Dterm}
V=\frac{1}{a}\left(\frac{3}{\mathcal{X}}-3\right)^{2}=\frac{9}{a}\left[e^{-\sqrt{2/3}\phi}-1 \right]^{2} 
\ee
where 
\be{X}
\mathcal{X}=e^{\sqrt{2/3}\phi}=\mathcal{T}+\mathcal{\bar{T}}-\frac{1}{3}\Phi
\ee
which corresponds to the Starobinsky's potential, as mentioned above
(all scalar fields are adimensionalized in Planck units).

The off-shell formulation of the minimal Starobinsly lagrangian during inflation
is determined by
\be{K}
\mathcal{K}=-3\log [\mathcal{T}+\mathcal{\bar{T}}-\Phi(z,\bar{z})],\,\,\,\mathcal{W}_{I}\rightarrow 0
\ee
The corresponding gravitino mass is 
\be{corr}
m_{\tilde{G}}=e^{\mathcal{K}/2}\frac{\mathcal{W}}{M_{Pl}^{2}}=e^{-\sqrt{\frac{3}{2}}\phi} \frac{\mathcal{W}}{M_{Pl}^{2}}\rightarrow 0
\ee
We shall now assume that supersymmetry is spontaneously broken at scales higher
than the inflation reheating, so that the superpotential $\mathcal{W}$ can be set to 
a constant $\mathcal{W}_{0}>0$.
As a consequence, a continuos spectrum of gravitinos  
will be produced during the inflation, with an average mass of 
\be{mass}
\langle m_{\tilde{G}}\rangle \simeq \langle e^{-\sqrt{\frac{3}{2}}\phi} \rangle_{\Delta N} \frac{\mathcal{W}_{0}}{M_{Pl}^{2}}\simeq 0.15\frac{\mathcal{W}_{0}}{M_{Pl}}
\ee
taking into account that the 
the inflationary plateau 
has a width of $\Delta \phi\simeq 5M_{Pl}$ corresponding to $\Delta N=\log a_{f}/a_{i}\simeq 60$ e-folds
of slow-roll inflation. 
A useful first approximation is to set $\langle \phi \rangle\simeq \Delta \phi/2$. 
In particular, $\phi(t_{R})\simeq M_{Pl}$ while $\phi(t_{R}-\Delta t)\simeq 6M_{Pl}$ 
with a $\Delta t$ time scale corresponding to $\Delta N$. 
As a consequence, Eq.(\ref{corr}) implies that a spectrum of massive gravitinos with $m_{\tilde{G}}\simeq 2\times (0.4\times 10^{-4} \div 1) \langle m_{\tilde{G}} \rangle$
is generated during slow roll. Fig.1 displays the precise Gravitino mass as a function of the inflaton field.

\subsection{Comments on the vacuum state with spontaneously broken 
 R-symmetry and SUSY}

As mentioned above, in Starobinsky's supergravity, the condition $\mathcal{W}_{I}\rightarrow 0$
during inflation is a viable way-out to the second modulus problem. 
The superpotential 
rolls down to zero before the inflation epoch. For $\mathcal{W}_{I}=0$, the vacuum 
state is R-symmetric and SUSY during the inflation stage.
This condition avoids any dangerous dynamics of the second modulus field,
potentially ruining conditions for a successful inflation. The condition $\mathcal{W}_{I}=0$
 implies a massless gravitino during inflation, which is incompatible with our suggestion. 
On the other hand, the spontaneous symmetry breaking of $U_{R}(1)$ and SUSY,
before or at least during inflation epoch while after the 
fast rolling down of the superpotential,
can only generate a constant contribution to the superpotential as
$\rightarrow \mathcal{W}_{0}={\rm  const}\neq 0$. 
This implies that the $\mathcal{G}$-term gets an extra contribution
\begin{equation}
\label{GKK}
\Delta\mathcal{G}=\log \mathcal{W}_{0}+\log \bar{\mathcal{W}}_{0}={\rm const}
\end{equation}
which implies a constant shift of the $V_{F}$-term
as 
\begin{equation}
\label{VF}
\Delta V_{F}=-3 \mathcal{W}_{0}\bar{\mathcal{W}}_{0}
\end{equation}
(only dependent by derivative of $\mathcal{G}$)
and a constant shift of the $V_{D}$-term as
$2\zeta\Delta V_{D}=-12$.
As a consequence, the inflaton potential 
is only shifted by a constant factor.
These numerical factors are not very important: they can be reabsorbed 
in the normalization of the Starobinsky's potential, as often discussed in literature.
So that, 
the spontaneous symmetry breaking 
of $U_{R}(1)$ and SUSY cannot contribute with dynamical interactions term 
to $\mathcal{W}_{I}$, i.e. it cannot destabilize the second modulus field. 
For instance, the R-symmetry implemented in Eqs.(\ref{V})
has fixed the structure of the potential Eq.(\ref{Dterm})
under the condition on $\mathcal{W}_{I}$. 
One can see that the only effect of a $\mathcal{W}_{0}={\rm  const}\neq 0$
during the inflation is the shift of the potential Eq.(\ref{Dterm}) of a constant factor
and the $z_{I}$ fields remain stabilized.

\begin{figure}[htb] \label{GMPLB}
\begin{center}
\caption{Gravitino mass function of Starobinsky inflaton. 
In the x-axis, the inflaton field is conveniently normalized in Planck units, 
while in the y-axis the gravitino mass function is normalized with respect of the average gravitinos mass
$\langle m_{\tilde{G}} \rangle$ (in $log_{10}$ scale in the $y$-axis). 
In particular, the oscillating epoch effectively starts at $\phi\//M_{P}\simeq 1$. 
On the other hand, the slow-roll effectively starts at $\phi/M_{P}\simeq 6$. 
 $\Delta\phi/M_{P}\sim 1\div 6$ is the gravitino production epoch.
So that, a continuos spectrum of super-heavy gravitinos is produced. }
\includegraphics[scale=0.06]{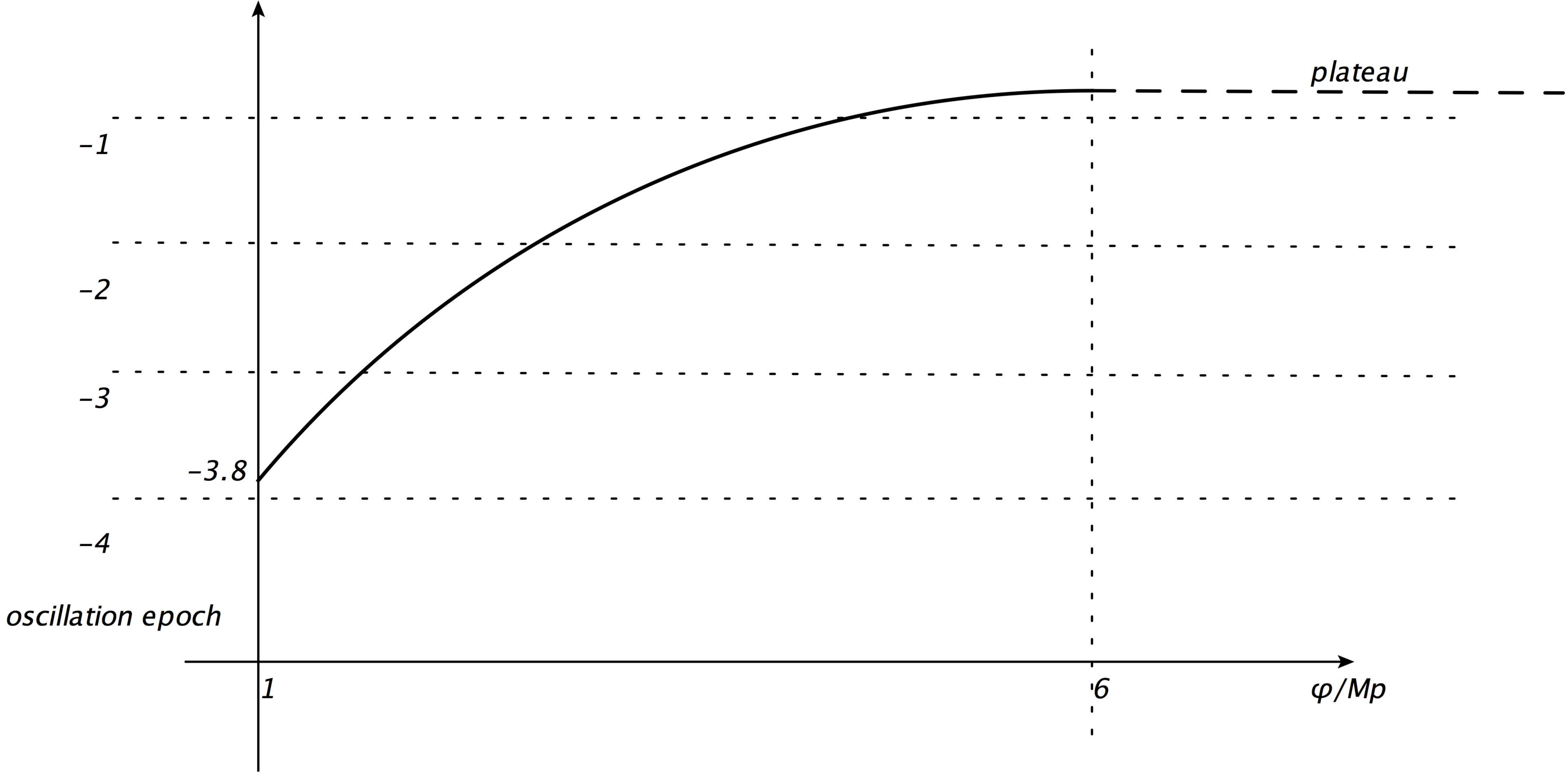}  
\end{center}
\vspace{-1mm}
\end{figure}

\section{Gravitino as SHDM}

In this section, we shall discuss how a correct Cold Dark Matter (CDM) abundance 
can be recovered in $R+\zeta R^{2}$ supergravity. 
In particular, we shall discuss the non-thermal production of 
gravitinos during inflation. 
The non-thermal Super Heavy Dark matter production triggered by inflation 
was studied in the simpler case of a scalar DM particle in \cite{Chung:1998zb}.
However, this mechanism can be implemented for 
gravitinos, even formally more subtle and never discussed in literature by other authors. 

First of all, the gravitino field in the full $R+\zeta R^{2}$ supergravity 
is described by  the Rarita-Schwinger action in presence the of a FRW dynamical metric:
\begin{equation}
\label{R}
S=\int d^{4}x e \bar{\psi}_{\mu}\mathcal{R}^{\mu}[\psi]
\end{equation}
\begin{equation}
\label{RS}
\mathcal{R}^{\mu}[\psi]=i \gamma^{\mu\nu\rho}\mathcal{D}_{\nu}\psi_{\rho}+m_{\tilde{G}}\gamma^{\mu\nu}\psi_{\nu}
\end{equation}
\begin{equation}
\label{cov}
\mathcal{D}_{\mu}\psi_{\nu}=\partial_{\mu}\psi_{\nu}+\frac{1}{4}\omega_{\mu ab}\gamma^{ab}\psi_{\nu}-\Gamma_{\mu\nu}^{\rho}\psi_{\rho}
\end{equation}
where $\gamma^{\mu_{1}...\mu_{n}}=\gamma^{[\mu_{1}}....\gamma^{\mu_{n}]}$,
$e={\bf det} e_{\mu}^{a}$  and $e_{a}^{\mu}$ is the (inverse) vielbein.
We can assume a torsion-free background metric so that $\Gamma_{\mu\nu}^{\rho}=\Gamma^{\rho}_{\nu\mu}$.

The EoM is
\begin{equation}
\label{eqRS}
(i  \mathcal{\slashed{D}}-m_{\tilde{G}})\psi_{\mu}-\left(i\mathcal{D}_{\mu}+\frac{m_{\tilde{G}}}{2}\gamma_{\mu}\right)\gamma \cdot \psi=0
\end{equation}

In a FRW cosmological background during inflation, 
Eq.(\ref{eqRS}) is reduced to 
\begin{equation}
\label{psiEOM}
i\gamma^{\mu\nu}\partial_{\mu}\psi_{\nu}=-\left(m_{\tilde{G}}+i\frac{a'}{a}\gamma^{0}\right)\gamma^{\mu}\psi_{\mu}
\end{equation}
with 
\begin{equation}
\label{cond}
e_{\mu}^{a}=a(\eta) \delta_{\mu}^{a},\,\,\,\,m_{\tilde{G}}=m_{\tilde{G}}(\eta),\,\,\,\, \omega_{\mu ab}=2\dot{a}a^{-1}e_{\mu[a}e_{b]}^{0}\,,
\end{equation}
and the solution reads 
\begin{equation}
\label{grav}
\psi_{\mu}(x)=\int \frac{d^{3}{\bf k}}{(2\pi)^{3}2k_{0}}\sum_{\lambda}\{ e^{i{\bf k}\cdot {\bf x}}b_{\mu}(\eta,\lambda)a_{k\lambda}(\eta)
\end{equation}
$$+e^{-i{\bf k}\cdot {\bf x}}b_{\mu}^{C}(\eta,\lambda)a_{k\lambda}^{\dagger}(\eta) \}$$

The corresponding mode equation is 
\begin{equation}
\label{mode}
\hat{P}^{\mu}b_{\mu}(\eta,\lambda)=0
\end{equation}
\begin{equation}
\label{P}
\hat{P}^{\nu}=i\gamma^{\mu \eta}\partial_{\eta}-\gamma^{\mu i}k_{i}-\left(m_{\tilde{G}}+i\frac{a'}{a}\gamma^{0}\right)\gamma^{\nu}
\end{equation}
Applying $\hat{P}^{\nu}$ on Eq.(\ref{mode}) 
as 
\begin{equation}
\label{mode2}
\hat{P}_{\nu}\hat{P}^{\mu}b_{\mu}(\eta,\lambda)=0
\end{equation}
we can rewrite the equation for modes in form 
\begin{equation}
\label{frequancy}
b_{\mu}''(\eta,\lambda)+\omega^{2}(k,a)b_{\mu}(\eta,\lambda)=0
\end{equation}
where $b''\equiv \partial^{2}b(\eta,\lambda)/\partial {\eta}^{2}\,$. 
The Eq.(\ref{mode}) can be rescaled as 
\begin{equation}
\label{frequancy}
b''_{\mu}(\hat{\eta},\lambda)+\omega^{2}(\hat{k},\hat{a})b_{\mu}(\hat{\eta},\lambda)=0
\end{equation}
where $\mu=m_{\tilde{G}}/H_{e}$, $\hat{\eta}/(a_{e}H_{e})=\eta$, $\hat{a}=a/a_{e}$
and $H_{e},a_{e}$ correspond to the oscillation epoch quantities 
- we have choosen this normalization for convenience.
The EoM can be solved imposing 
the boundary conditions. 
Let us comment that the fixing of boundary conditions corresponds to 
fixing the vacuum state. 
In order to calculate the number density of gravitinos 
produced, we perform 
the Bogoliubov transformation from the vacuum mode solution 
with boundary $\eta=\eta_{0}$
- corresponding to the initial cosmological time at which the vacuum state is 
specified- into the vacuum mode solution of boundary $\eta=\eta_{1}$
- corresponding to a generic later time at which gravitinos are no 
longer promoted from virtual to real particles (roughly we can assume $\phi \simeq M_{Pl}$ or so, close to the oscillation epoch). 
Let us note that the exact numerical values of $\eta_{0,1}$ 
are not important in the dynamical region 
$a'/a^{2}<<1$ or $\mu a/k<<1$. In this approximation, t
the EoM will be integrated with $\eta_{0}=-\infty$ and $\eta_{1}=+\infty$. 
We can define the Bogoliubov transformation as
\begin{equation}
\label{frequancy}
b_{\mu}^{(\eta_{1})}(\hat{\eta},\lambda)=\alpha_{k} b_{\mu}^{(\eta_{0})}(\eta,\lambda)+\beta_{k} b_{\mu}^{(\eta_{0})^{C}}(\eta)
\end{equation}
where $b_{\mu}^{(\eta_{1})}$ is the mode coefficient fixed on the Cauchy surface 
$\eta=\eta_{1}$ while $b_{\mu}^{(\eta_{0})}, b_{\mu}^{(\eta_{0})^{C}}$ are fixed on 
the Cauchy surface $\eta=\eta_{0}$,
where $\alpha_{k},\beta_{k}$ are the Bogoliubov's coefficient for 4-momenta $k$.
One can estimate the energy density of the gravitinos produced during inflation
as (see Appendix of Ref.\cite{Chung:1998zb} for a similar estimation)
\begin{equation}
\label{estimation}
\rho_{\tilde{G}}(\eta_{1})=\langle m_{\tilde{G}}\rangle n_{\tilde{G}}(\eta_{1})=\langle m_{\tilde{G}}\rangle H_{e}^{3}\left(\frac{1}{a(\eta_{1})} \right)\mathcal{P}
\end{equation}
with
\begin{equation}
\label{ps}
\mathcal{P}=\int_{0}^{\infty} \frac{d k}{2\pi^{2}}k^{2}|\beta_{k}|^{2}
\end{equation}
where we performed a Bogoliubov transformation from the Cauchy surface foliated by $\eta=\eta_{0}$
to another Cauchy surface with cosmological frame time 
$\eta_{1}>\eta_{0}$ and assuming the inflation conditions
$\dot{a}/a^{2}<<1$. As mentioned above, for formal convenience, we have normalized
$k\rightarrow k/a H_{e}$, $\eta\rightarrow \eta /a_{e}H_{e}$,
$a\rightarrow a/a_{e}$, where $e$
labels variables of the oscillation epoch. 
As usual, in such a procedure there is an apparent ambiguity in 
the definition of the vacuum. As mentioned above, such a problem is equivalent to 
the definition of the boundary conditions for Eq.(\ref{psiEOM}).
We remind that a systematical method of classification of the 
inequivalent vacuum states was suggested in Refs.\cite{Adiabatic1,Adiabatic2,Adiabatic3},
introducing the concept of adiabatic vacuum. 
From such a definition, it is possible to construct 
a set of solutions for the EoM (\ref{psiEOM})
reduced to the usual plane waves ($a'(\eta)=0$, for all $\eta$ values). 
Let us define the n-th adiabatic vacuum at a certain time $\eta^{*}$
by following boundary conditions:
\begin{equation}
\label{estimation}
b_{\mu}(\eta^{(n)})=b_{\mu}^{(n)}(\eta^{*}),\,\,\,\,b'_{\mu}(\eta^{*})=b'^{(n)}_{\mu}(\eta^{*})
\end{equation}
where $b_{\mu}^{(n)}(\eta)$ is a n-th order perturbative expansion of the complete solution, satisfying the n-th adiabatic order in the asymptotic limit 
(see Ref.\cite{20} for a general and more detailed definition).

We estimate the relation among the gravitino energy density 
normalized over the radiation as: 
\begin{equation}
\label{gravitino}
\frac{\rho_{\tilde{G}}(t_{0})}{\rho_{R}(t_{0})}=\frac{\rho_{\tilde{G}}(t_{R_{e}})}{\rho_{R}(t_{R_{e}})}\left(\frac{T_{R}}{T_{e}} \right)
\end{equation}
where 
$\rho_{\tilde{G}}(t_{Re})/\rho_{R}(t_{Re})$
is determined after the Reheating epoch,
 and $t_{0}$ is the present cosmological time.
Gravitinos were produced during the $t_{e}>t_{Rh}$
epoch, i.e. during the inflaton oscillations 
and decays into Susy SM (or Beyond SM) particles.
$\rho_{\tilde{G}}(t_{Re})/\rho_{R}(t_{Re})$
is estimated as
\begin{equation}
\label{XR}
\frac{\rho_{\tilde{G}}(t_{Rh})}{\rho_{R}(t_{Rh})}\simeq \frac{8\pi}{3}\left(\frac{\rho_{\tilde{G}}(t_{e})}{M_{Pl}^{2}H^{2}(t_{e})} \right)
\end{equation}
The inflaton mass is the characteristic scale for the Hubble constant calculated in $t_{e}$:
$H^{2}(t_{e})\sim m_{\phi}^{2}$ and $\rho(t_{e})\sim m_{\phi}^{2}M_{Pl}^{2}$.
and this implies the following relations for gravitino abundance 
\begin{equation}
\label{begine}
\Omega_{\tilde{G}}h^{2}\sim 10^{17}\, \left( \frac{T_{Rh}}{10^{9}\, \rm GeV}\right)\left(\frac{\rho_{\tilde{G}}(t_{e})}{\rho_{c}(t_{e})} \right)
\end{equation}
where $\rho_{c}(t_{e})=3H(t_{e})^{2}M_{Pl}^{2}/8\pi$ is the critical energy density during $t_{e}$.
Eq.(\ref{begine}) can be conveniently rewritten as 
\begin{equation}
\label{rewe}
\Omega_{\tilde{G}}h^{2}\simeq \Omega_{R}h^{2}\left(\frac{T_{Rh}}{T_{0}} \right)\frac{8\pi}{3}\left( \frac{\langle m_{\tilde{G}}\rangle }{M_{Pl}}\right)\frac{n_{\tilde{G}}(t_{e})}{M_{Pl}H^{2}(t_{e})}
\end{equation}
As explicitly shown in Eq.(\ref{rewe}), the gravitino mass is up to the inflaton mass and the reheating temperature. 
However, the inflaton mass is constrained to be 
 $m_{\phi} \simeq 10^{13}\, \rm GeV$ or so. 
 On the the other hand
$T_{Rh}/T_{0}\simeq 4.2\times 10^{14}$
for a successful reheating. 
As a consequence, a correct abundance of cold dark matter
can be recovered 
for a gravitino mass of
$\langle m_{\tilde{G}}\rangle \simeq (10^{-2}\div 1)\times m_{\phi}$
$\simeq 10^{11}\div 10^{13}\, \rm GeV$, 
constraining $W_{0}$ in Eq.(\ref{mass}). 
As a result the SUSY symmetry breaking scale
is expected to be around the gravitino mass.  
In particular, all other superparticles are assumed to be heavier than the gravitino.

\section{Comments on string non-perturbative contributions}

Our model could be UV completed in context of string theory.
It is commonly retained that in the limit of $\alpha'=l_{s}^{2}\rightarrow 0$, 
 superstrings reduce to supergravity models. 
However non-perturbative stringy corrections can 
generate new effective superpotentials
which are not allowed at perturbative level.  
In our framework, stringy corrections 
can destabilize the gravitino leading to possible phenomenological 
implications for indirect detection of dark matter. 
In particular, the initial $U(1)_{R}$ gauge symmetry can 
be broken
by exotic stringy instantons, 
i.e. by Euclidean D-brane instantons 
 of open superstring theories 
or worldsheet instantons in heterotic superstring theory
(See \cite{Bianchi:2009ij} for a review on this subject). 
For example, the generation of 
$\mu HL$ superpotentials by 
 $E2$-branes 
 in intersecting D6-brane models 
 was discussed in Ref. \cite{Addazi:2015fua}.
 The associated effective lagrangian is 
\be{LHL}
\mathcal{L}_{E2}=C^{(1)}\beta^{(1)}H_{u_{A}}\tau_{A}^{(1)}+C_{i}^{'(1)}\gamma^{(1)}L_{A}^{i}\tau_{A}^{(1)}
\ee
where $\beta^{(1)},\gamma^{(1)},\tau^{(1)}$ are fermionic zero modes, 
which correspond to excitations of open strings attached to 
$U(1)-E2$, $U(1)'-E2$ and $Sp_{L}(2)-E2$ respectively. 
Integrating out fermionic zero modes, 
one obtains 
\be{W11}
\int d^{2}\theta\, \mathcal{W}=\int d^{2}\theta \int d^{2}\tau^{(1)}d\beta^{(1)}d\gamma^{(1)}e^{\mathcal{L}_{E}}
\ee
$$=M_{S}e^{-S_{E2}}(C^{(1)}C_{i}^{'(1)})H_{u}L^{i}$$
where $M_{S}$ is the string scale and
$e^{-S_{E2}}$
 is controlled by the geometric scalar moduli which parametrize the 
  3-cycles, wrapped by the $E2$-instanton on the Calabi-Yau $CY_{3}$.  

On the other hand, in NMSSM scenarios, the introduction of 
a chiral singlet superfield $\mathcal{S}_{R}$ can allow the non-perturbative generation
of suppressed effective superpotential of the type
\be{OPER}
\mathcal{W}_{R}=\mu_{i}\left(\frac{S_{R}}{M_{S}}\right)^{n}H_{u}L^{i}
\ee

The first term of Eq.(\ref{W11}) is dangerous, since 
 the gravitino has also a coupling with
$W^{\pm},Z,\gamma,V_{R}$ gauge bosons and their related gauginos of the form 
\be{gravitino}
L_{int}=-\frac{i}{8M_{Pl}}\bar{\psi}_{\mu}[\gamma^{\nu},\gamma^{\rho}]\gamma^{\mu}\lambda F_{\nu\rho}
\ee
As usual, neutral gauginos mix with higgsinos, and their mass eigenstates are neutralinos. 
So that, from (\ref{W11}) and (\ref{gravitino}), neutralinos mediate two-body decays
$\tilde{G}\rightarrow \gamma\nu, Z\nu, V_{R}\nu$. 
In particular $\tilde{G}\rightarrow \gamma \nu$ is the easier decay to constrain since 
very high energy gamma rays and neutrinos with a peak distribution 
are produced. The associated decay rate is 
\be{Gamma}
\Gamma^{(0)}_{\tilde{G}\rightarrow \gamma \nu}=\frac{1}{32\pi}\cos^{2}\theta_{W}\frac{m_{\nu}}{m_{\chi}}\frac{m_{\tilde{G}}^{3}}{M_{Pl}^{2}}\left(1-\frac{m_{\nu}^{2}}{m_{\tilde{G}}^{2}} \right)^{3}\left(1+\frac{m_{\nu}^{2}}{3 m_{\tilde{G}}^{2}} \right)
\ee
Now, in our high scale supersymmetry breaking, 
assuming 
$m_{\chi}\simeq 10^{13}\, \rm GeV$
and $m_{\tilde{G}}\simeq 10^{11}\, \rm GeV$, 
the decay rate is of only 
$\Gamma^{0}\simeq 10^{-20}\, \rm eV$
corresponding to $\tau^{0} \simeq 10^{5}\, \rm s$. 
This implies that non-perturbative stringy instantons 
generating the operator (\ref{W11})
can 
be very dangerous: they completely destabilize 
gravitino Dark Matter 
and they have to be suppressed in non-perturbative regime. 
This is possible if specific non-perturbative RR or NS-NS fluxes are wrapped by 
the instantonic Euclidean D-brane \cite{Bianchi:2012pn}.
Calling $\mathcal{N}_{N.P.}$ the non-pertubative suppression factor, 
this can screen the the bare decay rate as 
$\Gamma=\mathcal{N}_{N.P.}\Gamma_{0}$. 
A suppression factor $\mathcal{N}\simeq 10^{-11}$
in order to get a gravitino cosmological life-time of at least $1\, \rm Gyr$ or so. 

On the contrary, operator like (\ref{OPER}) can destabilize the gravitinos 
with an overall suppression $(\langle \phi_{S}\rangle/M_{S})^{n}$
assuming that the singlet gets a vacuum expectation value. 
The corresponding decay time has to be suppressed up to a cosmological time scale $\tau=\left(M_{S}/\langle\phi_{S}\rangle\right)^{n}\tau^{0}>1\, \rm Gyr$. 
For $n=1$, $\langle \phi_{S} \rangle\simeq 10^{-11}M_{S}$ saturates the bound. 
Assuming $M_{S}=g_{S}M_{Pl}\simeq 10^{16}\, \rm GeV$, the scalar singet mass
is around $100\, \rm TeV$, which could be reached by the next generation of colliders, 
with decay channels strongly depending on the completion of our model.
On the other hand, for $n>1$ the scalar singlet field is heavier than $100\, \rm TeV$.  
This opens the interesting possibility of super-heavy gravitino decays 
$\tilde{G}\rightarrow \gamma \nu$ with two photons  and neutrino peaks 
of energy $E_{CM}\simeq m_{\tilde{G}}/2\simeq 10^{8}\div 10^{13}\, \rm GeV$.

The observation of a so high energy neutrinos and photons could be a strong indirect evidence in 
favor of our scenario. In particular, these very high energy neutrinos can be observed by 
AUGER, Telescope Array, ANTARES and IceCube. 
and while eventually they could not be explained by any possible astrophysics sources. 

\section{Conclusions and discussions}

In this paper, we have discussed some implications of a $R+\zeta R^{2}$ supergravity model 
with supersymmetry broken at high scales. 
As is well known, the Starobinsky (super)gravity is in agreement with Planck data. 
Then, we showed how
 this model can also 
provide a good candidate of Super Heavy Gravitino Dark Matter.
Gravitinos can be non-thermally produced during inflationary slow-roll. 
Intriguingly, in the spaces of parameter of  inflaton field and of this gravitino are connected. 
This model provides a new peculiar prediction: Super-Heavy Gravitinos are produced with a continuos mass spectrum, 
following the inflaton field. In our framework, 
CDM data can be constrained ny
the inflaton potential (and viceversa). 
Finally, we commented on possible problems in the UV completion of our supergravity model 
in contest of superstring theories. In particular, even if the gravitino can be protected by R-parity 
in perturbative supergravity, it will be not protected by any custodial discrete or abelian gauge symmetries in non-perturbative strings regime. 
The gravitino can be destabilized very fast, even in the limit of $\alpha'\rightarrow 0$. 
In addition, the famous problem of string moduli stabilization during inflation is still present. 
But non-perturbative effects 
can strongly suppress a certain class of operators generated by Euclidean D-branes of worldsheet instantons. 
It is conceivable that the non-perturbative UV protection cannot avoid all possible R-parity violating 
gravitino decays. This implies that the gravitino can decay in a cosmological time 
in several channels. In particular two-body decays $\tilde{G}\rightarrow \gamma \nu$ can produce
very high energy peaks of neutrinos and photons, 
of $E_{CM}\simeq 10^{3}\div 10^{7}\,PeV$. The detection of these very high energy neutrinos 
with a peak-like two-body decay distribution could be a strong indirect hint for our model. 
On the other hand, we also relate our proposal with the presence of a new scalar singlet 
at $100\, \rm TeV$, which could be detected at future high energy colliders. 

Our suggestion should extend to a more general class of $f(R)$-supergravity, 
like $R+R^{n}$, with $n>2$, studied in \cite{Ferrara:2013pla}. 
Many attempts  to unified phantom dark energy and inflation were suggested in 
contest of $f(R)$-gravity 
\cite{Nojiri:2005pu,Capozziello:2005tf} (see \cite{Capozziello:2011et} for a review on general aspects of $f(R)$-gravity). The UV completion of more general $f(R)$-gravity models 
can provide a unifying picture of dark matter, dark energy and inflation \footnote{The same attitude has inspired 
recent papers suggesting a unifying picture of dark matter and dark energy from 
a hidden strong sector \cite{Addazi:2016sot,Addazi:2016nok}. On the other hand, 
possible implications in direct detection of similar models in the framework of hidden asymmetric standard model
were recently discussed in Ref.\cite{Addazi:2015cua}.} \footnote{We mention that in contest of more generic 
$f(R)$-gravity, it remains still unclear the role of primordial black holes, which 
apparently have a antievaporation instability rendering impossible their evaporation \cite{Addazi:2016prb}.}.
The same mechanism of gravitinos production 
discussed in Section III could also be implemented in string-inspired climbing scalar pre-inflationary models \cite{Dudas:2012vv,Sagnotti:2013ica,Kitazawa:2014dya}.
In this case, a more complicated mass density spectrum of gravitinos is expected and
pre-inflationary produced gravitinos should be expected to be part of the CDM composition. 
However, a detailed analysis deserves a separate analysis beyond the purpose of this letter. 
Finally, we mention that the parameters space of gravitinos mass can change if a consistent amount 
of Primordial Black Holes   \cite{Khlopov:1985jw,Khlopov:2004tn,Khlopov:2008qy} were produced during the early Universe. In this case, superheavy gravitinos could have been produced out of thermal equilibrium after the reheating 
by PBHs evaporation.

\vspace{0.5cm}

{\large \bf Acknowledgments}
\vspace{3mm}

AA would like to thank M. Bianchi, S. Capozziello, A. Marciano, S. Odintsov, M.~Porrati, A. Sagnotti and the anonymous referee for 
useful discussions and remarks. 
The work of AA was supported in part by the MIUR research grant Theoretical Astroparticle Physics PRIN 2012CP-PYP7 and by SdC Progetto speciale Multiasse La Societ\`a  della Conoscenza in Abruzzo PO FSE Abruzzo 2007-2013. 
The work by MK was performed within the framework of the Center FRPP supported by MEPhI Academic Excellence Project (contract 02.03.21.0005, 27.08.2013), in which the part on initial cosmological conditions was supported by the Ministry of Education and Science of Russian Federation,
project 3.472.2014/K  and on the forms of dark matter by grant RFBR 14-22-03048.

\end{document}